\documentclass[preprint2,10.5pt]{aastex}
\begin{document}

\title{\large{\rm{DEEP INFRARED ZAMS FITS TO BENCHMARK OPEN CLUSTERS \\ HOSTING $\delta$ SCUTI STARS}}}
\author{Daniel J. Majaess$^{1,2}$, David G. Turner$^{1,2}$, David J. Lane$^{1,2}$, Tom Krajci$^{3,4}$}
\affil{$^1$ Saint Mary's University, Halifax, Nova Scotia, Canada}
\affil{$^2$ The Abbey Ridge Observatory, Stillwater Lake, Nova Scotia, Canada}
\affil{$^3$ Astrokolkhoz Telescope Facility, Cloudcroft, New Mexico, USA}
\affil{$^4$ American Association of Variable Star Observers, Cambridge, MA, USA}
\email{\rm{dmajaess@cygnus.smu.ca}}

\begin{abstract}
This research aims to secure precise distances for cluster $\delta$ Scutis in order to investigate their properties via a $VI$ Wesenheit framework. Deep \textit{JHK$_s$} colour-colour and ZAMS relations derived from $\simeq700$ unreddened stars featuring 2MASS photometry and precise Hipparcos parallaxes ($d\lesssim25$ pc) are applied to establish distances to several benchmark open clusters that host $\delta$ Scutis: Hyades, Pleiades, Praesepe, $\alpha$ Persei, and M67 ($d=47\pm2,138\pm6,183\pm8,171\pm8,815\pm40$ pc).  That analysis provided constraints on the $\delta$ Sct sample's absolute Wesenheit magnitudes ($W_{VI,0}$), evolutionary status, and pulsation modes (order, $n$).  The reliability of $JHK_s$ established cluster parameters is demonstrated via a comparison with \citet{vl09} revised Hipparcos results.  Distances for 7 of 9 nearby ($d\le250$ pc) clusters agree, and the discrepant cases (Pleiades \& Blanco 1) are unrelated to (insignificant) $T_{e}-(J-K_s)$ variations with cluster age or iron abundance.  $JHK_s$ photometry is tabulated for $\simeq3\times10^3$ probable cluster members on the basis of proper motions (NOMAD).  The deep $JHK_s$ photometry extends into the low mass regime ($\simeq0.4 M_{\sun}$) and ensures precise ($\le5$\%) ZAMS fits.  Pulsation modes inferred for the cluster $\delta$ Scutis from $VI$ Wesenheit and independent analyses are comparable ($\pm n$), and the methods are consistent in identifying higher order pulsators.  Most small-amplitude cluster $\delta$ Scutis lie on $VI$ Wesenheit loci characterizing $n\ge1$ pulsators.  A distance established to NGC 1817 from $\delta$ Scutis ($d\simeq1.7$ kpc) via a universal $VI$ Wesenheit template agrees with estimates in the literature, assuming the variables delineate the $n\ge1$ boundary. Small statistics in tandem with other factors presently encumber the use of mmag $\delta$ Scutis as viable distance indicators to intermediate-age open clusters, yet a $VI$ Wesenheit approach is a pertinent means for studying $\delta$ Scutis in harmony with other methods. 
\end{abstract} 

\section{{\rm \footnotesize INTRODUCTION}}

$\delta$ Sct variables are unique among standard candles of the classical instability strip for permitting the determination of distances to population I and II environments from a single $VI$ Wesenheit calibration.  SX Phe and other metal-poor population II $\delta$ Scutis lie toward the short-period extension of the Wesenheit ridge characterizing population I $\delta$ Scutis \citep[Fig.~3 in][see also \citealt{ph98}]{ma10c}.  That presents an opportunity to bridge and strengthen the distance scales tied to globular clusters, open clusters, nearby galaxies, and the Galactic center where such variables are observed \citep[][]{mc00,mc07,po10,ma10c}.  To that end the present research examines $\delta$ Sct calibrators associated with benchmark open clusters via a $VI$ Wesenheit framework, an analysis which relies on the establishment of precise cluster distances and multiband photometry ($VIJHK_s$). 

In this study, infrared colour-colour and ZAMS relations are constructed from unreddened stars in close proximity to the Sun with precise Hipparcos parallaxes (\S \ref{zams}).  The relations are subsequently employed to establish parameters for five benchmark open clusters which host $\delta$ Scutis, namely the Hyades, Pleiades, Praesepe, $\alpha$ Persei, and M67 (\S \ref{zams}).  Cluster membership provides constraints on the absolute Wesenheit magnitudes ($W_{VI,0}$), evolutionary status, and pulsation modes (order, $n$) for the $\delta$ Scutis (\S \ref{sdsct}).  An independent determination of the cluster parameters is pursued since the objects form the foundation of the open cluster scale and yet their distances are uncertain.  Most notably the Hipparcos distance to stars in the Pleiades is $d=120.2\pm1.9$ pc \citep{vl09}, whereas HST observations imply $d=134.6\pm3.1$ pc \citep{so05}.  Likewise, four Hipparcos-based distances cited in the literature for $\alpha$ Persei disagree (Table~\ref{table1}). In \S \ref{sdsct}, a Wesenheit analysis ($VI$) is shown to be a viable means for investigating $\delta$ Scutis. Lastly, the distance to NGC 1817 is evaluated via a universal Wesenheit template using new $VI$ photometry acquired from the Abbey Ridge Observatory \citep[ARO,][]{la07,tu09} for the cluster's numerous $\delta$ Scutis (\S \ref{sngc1817}).

\section{{\rm \footnotesize $JHK_s$ INTRINSIC RELATIONS}}
\label{zams}
\begin{figure}[!t]
\begin{center}
\includegraphics[width=7cm]{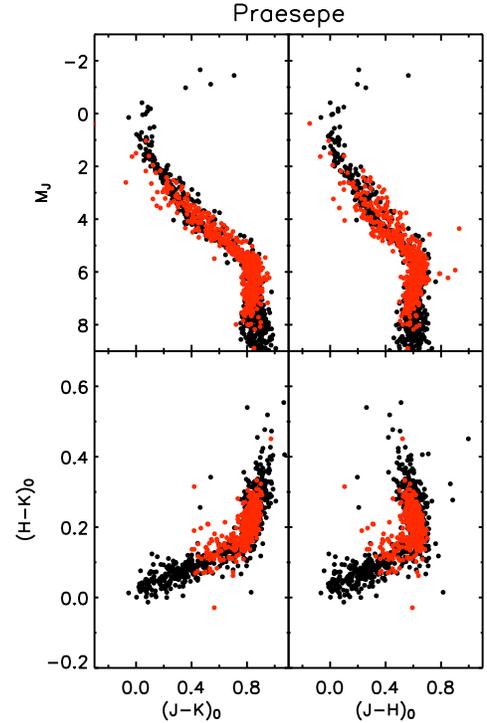} 
\caption{\small{Deep 2MASS $JHK_s$ colour-magnitude and colour-colour diagrams for the calibration (red dots) and Praesepe star cluster (black dots). Likely members of the Praesepe were selected on the basis of proper motions (NOMAD).  The resulting parameters are $E(J-K_s)=0.025\pm0.015$ and $d=183\pm8$ pc.}}
\label{fig1}
\end{center}
\end{figure}

Intrinsic colour-colour and ZAMS relations are derived from $JHK_s$ photometry for A, F, G, K, and M-type stars that feature precise parallaxes ($d\lesssim25$ pc).  The infrared photometry and parallaxes are provided via the 2MASS and Hipparcos surveys \citep{pe97,cu03,sk06}.   2MASS photometry may be saturated for nearby stars, yet reliable data are available for fainter stars of late spectral type.  The photometric uncertainties thus increase for brighter early-type stars (Fig.~\ref{fig1}), which are already undersampled owing to the nature of the initial mass function.   Spurious data deviating from the evident intrinsic functions were excluded. 

$M_J$/$(J-H)_0$ and $M_J$/$(J-K_s)_0$ colour-magnitude diagrams for the calibration are presented in Fig.~\ref{fig1} (red dots).  The latter passband combination exhibits smaller uncertainties.  $(J-H)_0$/$(H-K_s)_0$ and $(J-K_s)_0$/$(H-K_s)_0$ colour-colour diagrams are likewise shown in Fig.~\ref{fig1} (red dots).  Infrared relations offer advantages over those founded on $UBV$ photometry owing to the mitigated impact of chemical composition (\S \ref{smetallicity}, Fig.~\ref{fig2}), differential reddening, and total extinction.  

Colour-magnitude and colour-colour diagrams were assembled for the target clusters by obtaining 2MASS photometry for members on the basis of proper motions (Figs.~\ref{fig1},~\ref{fig2b}).  The Naval Observatory Merged Astrometric Dataset \citep[NOMAD,][]{za04} features proper motion data for the fields inspected.  $JHK_s$ photometry for $\sim3\times10^3$ probable clusters members were tabulated, and the list includes several previously unidentified members.  That sample shall be made available online via the \textit{Centre de Donn\'ees astronomiques de Strasbourg}.  Photometry for the brighter stars may be saturated, as noted earlier, while lower mass members are catalogued despite being near the faint-limit of the 2MASS survey.  High-precision multi-epoch $JHK_s$ data for M67 is available from \citet[][see also \citealt{sa09}]{ni00}.

%---------------------------  FIGURE 1 --------------------------%
%---------------------------  end FIGURE 1 --------------------------%

%--------------------  TABLE 1 ---------------------- %

\begin{table*}[t]
\centering
\begin{minipage}{17.2cm}
\caption[]{Distances to Benchmark Open Clusters}
\label{table1}
\begin{tabular}{lllllll}
\hline 
\hline \rule{0pt}{2.2ex}
Cluster & HIP (M97) & HIP (R99) & HIP (V99) & HIP (V09) & $JHK_s$ & HST \\ 
\hline \rule{-2.5pt}{2.3ex}
M67 & $\cdots$ & $\cdots$ & $\cdots$ &$\cdots$ & $815\pm40$ pc & $\cdots$   \\
Hyades &  $\cdots$ & $\cdots$ & $\cdots$ & $46.45\pm0.50$ pc & $47\pm2$ pc & $48.3\pm2.0$ pc \\
Praesepe & $177.0\pm10.3$ pc & $180.5\pm10.7$ pc  & $180$ pc & $181.6\pm6.0$ pc & $183\pm8$ pc &  $\cdots$  \\
Pleiades & $116.3\pm3.3$ pc & $118.2\pm3.2$ pc & $125$ pc & $120.2\pm1.9$ pc & $138\pm6$ pc & $134.6\pm3.1$ pc \\
$\alpha$ Persei & $184.2\pm7.8$ pc & \textbf{190.5}$\pm7.2$ pc & $170$ pc & \textbf{172.4}$\pm2.7$ pc & $171\pm8$ pc & $\cdots$   \\ 
\hline \rule{-2.5pt}{2.3ex}
Coma Ber* & $88.2\pm1.7$ pc  & $87.0\pm1.6$ pc & $86$ pc &$86.7\pm0.9$ pc & $85\pm6$ pc & $\cdots$   \\
Blanco 1 & $252.5\pm31.1$ pc & \textbf{263}$\pm31$ pc & $190$ pc & \textbf{207}$\pm12$ pc & $240\pm10$ pc & $\cdots$   \\
IC 2391* & $147.5\pm5.4$ pc & $146.0\pm4.7$ pc & $140$ pc &$144.9\pm2.5$ pc & $134\pm13$ pc & $\cdots$   \\
IC 2602* & $146.8\pm4.7$ pc & $152\pm3.7$ pc & $155$ pc &$148.6\pm2.0$ pc & $147\pm14$ pc & $\cdots$   \\
NGC 2451* & $\cdots$  & $188.7\pm6.8$ pc & $220$ pc &$183.5\pm3.7$ pc & $189\pm15$ pc & $\cdots$   \\
\hline \\
\end{tabular}
\small{Notes: Hipparcos (HIP) distances from \citet[M97]{me97}, \citet[R99]{ro99}, \citet[V99]{vl99}, and \citet[V09]{vl09}. HST distances to the Hyades and Pleiades from \citet{va97} and \citet{so05}. $JHK_s$ results from this study.   Clusters highlighted by an asterisk were more difficult to assess ($JHK_s$).  Iron abundances and age estimates for the clusters are tabulated in \citet{me97} and \citet{vl99}.}
\end{minipage}
\end{table*}
%--------------------  TABLE 1 ---------------------- %
Reddenings for the target clusters were secured by shifting the intrinsic colour-colour relations to the observed data (Fig.~\ref{fig1}).  $JHK_s$ extinction laws were adopted from \citet{bona08} and references therein.  The distance to a cluster follows by matching the ZAMS to the observed data for the reddening established by the intrinsic colour-colour relations.  Precise results were obtained because the trends for late-type stars in $JHK_s$ colour-magnitude and colour-colour diagrams provide excellent anchor points for fitting ZAMS and intrinsic relations (Figs.~\ref{fig1},~\ref{fig2b}). $J-K_s$ and $J-H$ were observed to remain nearly constant and become bluer with increasing magnitude ($M_J\gtrsim6$) for low mass main-sequence stars beginning near spectral type M (Figs.~\ref{fig1},~\ref{fig2b}, see also \citealt{sa09} and references therein), and a sizable separation exists between main-sequence and evolved M-type stars in the $JHK_s$ colour-colour diagram \citep[see also][]{sl09}.  \citet{tu10b} developed intrinsic $JHK_s$ functions to describe early-type stars ($\lesssim$K0) via an alternate approach.  

Distances obtained for the benchmark clusters examined are summarized in Table~\ref{table1}.   

\subsection{{\rm \footnotesize AGE AND METALLICITY DEPENDENCIES}}
\label{smetallicity}
The $JHK_s$ distances agree with \citet{vl09} Hipparcos results for 7 of 9 star clusters within 250 pc (Table~\ref{table1}).  The distance determined here to the Pleiades favours the HST estimate rather than that established by Hipparcos (Table~\ref{table1}).\footnote{The HST and Hipparcos Pleiades surveys lack overlap, and the former exhibits comparatively smaller statistics.}  \citet{so05} (\& others) argue that the Hipparcos distance to the Pleiades is erroneous. Conversely, the reliability of the ZAMS distance to the Pleiades has been questioned owing to the possible neglect of colour-$T_{e}$ variations with stellar age and chemical composition.  The matter is now investigated.  

An age-luminosity effect has been proposed as a possible source for the disagreement between the ZAMS and Hipparcos distance to the Pleiades.  That hypothesis is not supported by the infrared distances established here.  The Hyades, Praesepe, and $\alpha$ Persei clusters bracket the discrepant cases of the Pleiades and Blanco 1 in age (Table~\ref{table1}).   $JHK_s$ and \citet{vl09} Hipparcos distances to the former clusters are in broad agreement (Table~\ref{table1}).  The comparatively nearby ($d\le250$ pc) open clusters IC 2602, NGC 2451, IC 2391, and Coma Ber were investigated to bolster the case, but proved  more difficult to assess.  The Hipparcos distance to the Pleiades implies that the cluster's ZAMS is a sizable $\simeq0^{\rm m}.4$ below the faintest calibration stars (Fig.~\ref{fig1}, where the photometric uncertainties are minimized, $M_J/(J-K_s)_0$). The Hipparcos zero-point for the Pleiades is inconsistent with a $M_J/(J-K_s)_0$ ZAMS calibration (Fig.~\ref{fig1}) which features stars of differing age, chemical composition, and peculiarities.

%---------------------------  FIGURE 2b --------------------------%
\begin{figure*}[!t]
\begin{center}
\includegraphics[width=13cm]{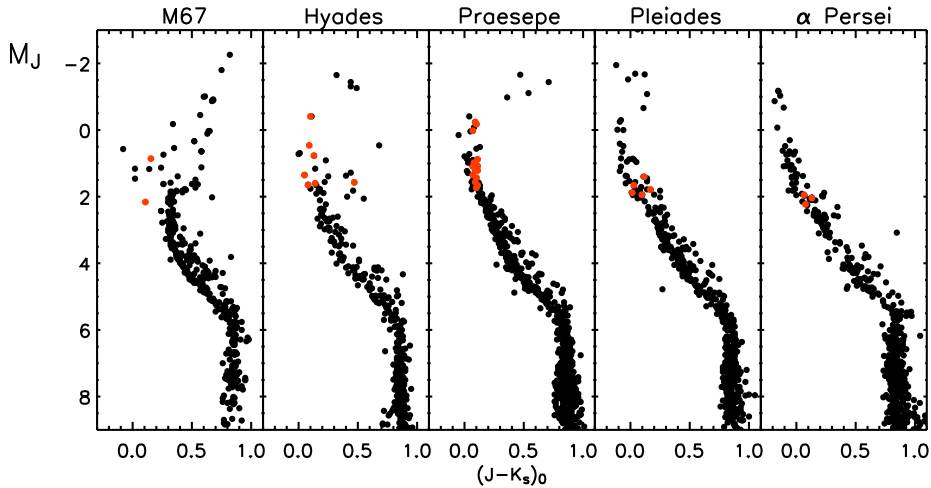} 
\caption{\small{Deep 2MASS $JHK_s$ colour-magnitude diagrams for M67, Hyades, Praesepe, Pleiades, and $\alpha$ Persei star clusters.  The clusters feature a common ZAMS morphology in the infrared.  The $\delta$ Scutis examined in \S \ref{sdsct} (red dots) may be evolved (Hyades/Praesepe), blue stragglers (M67), or occupy the binary / rapid rotator sequence (Pleiades).}}
\label{fig2b}
\end{center}
\end{figure*}
%---------------------------  end FIGURE 2b --------------------------%

The clusters in Table~\ref{table1} exhibiting discrepant distances are not correlated with iron abundance.  That supports \citet{al96} and \citet{pe05} assertion that $J-K_s$ is relatively insensitive to metallicity over the baseline examined.  \citet{pe05} suggested that $J-K_s$ may exhibit a marginal dependence on metallicity, but cautioned that the errors are sizable and the correlation coefficient is consistent with zero.  The impact of a marginal $T_{e}-{\rm [Fe/H]}-(J-K_s)$ dependence appears insignificant since stars belonging to the clusters examined display near solar iron abundances \citep[][their Table~1]{me97,vl99}.  270 calibration stars (Fig.~\ref{fig1}) featured in \citet{sou10} PASTEL catalogue of stellar atmospheric parameters exhibit a peak distribution near ${\rm [Fe/H]}\simeq-0.05$, which is analogous to or inappreciably less than members of the Pleiades (${\rm [Fe/H]}=-0.039\pm0.014,+0.03\pm0.05,+0.06\pm0.01$, \citealt{ta08,sod09,pa10}).  Colours for calibrating stars (Fig.~\ref{fig1}) in PASTEL were plotted as a function of effective temperature and metallicity (Fig.~\ref{fig2}, optical photometry from \citealt{me91}).  $B-V$ and $U-B$ colour indices appear sensitive to iron abundance whereby metal-rich stars are hotter at a given colour (Fig.~\ref{fig2}), as noted previously \citep[][and references therein]{tu79}.  Conversely, $J-K_s$ appears comparatively insensitive to iron abundance over the restricted baseline examined (Fig.~\ref{fig2}). Yet the results should be interpreted cautiously and in tandem with the other evidence presented given the semi-empirical nature of that analysis ($T_{e}$ and [Fe/H] are model dependent).  

The clusters exhibit a similar ZAMS morphology in the infrared ($M_J/(J-K_s)_0$, Figs.~\ref{fig1},~\ref{fig2b}).  The Hyades, Praesepe, Pleiades, and M67 ZAMS (unevolved members) are nearly indistinguishable (Fig.~\ref{fig2b}).  \citet{st03} likewise noted that the Praesepe and Pleiades cluster share a ZAMS ($M_v/(V-I)_0$) that is essentially coincident throughout.  By contrast, the apparent sensitivity of optical photometry to metallicity may explain (in part) certain anomalies which distinguish individual clusters in optical colour-magnitude and colour-colour diagrams \citep[Fig.~\ref{fig2}, see also][]{tu79,me97,st03,vl09}. Compounded uncertainties prevent a direct assessment of the infrared colour-colour function's universality ($(J-K_s)_0/(H-K_s)_0$).  Minimizing the uncertainties associated with the $JHK_s$ photometry and extending the restricted temperature baseline are desirable.  Fainter $JHK_s$ observations could be acquired from l'Observatoire Mont-M\'{e}gantique or the forthcoming $VVV$ survey \citep{ar09,ar10,mi10}, whereas brighter stars could be observed as part of the AAVSO's IR photoelectric photometry program \citep{he02,te09}.  

\textit{In sum}, neither variations in iron abundance or stellar age readily explain the discrepancies between the $JHK_s$ ZAMS and Hipparcos distances (Table~\ref{table1}).  It is emphasized that the problematic cases (the Pleiades \& Blanco 1) constitute the minority (Table~\ref{table1}).  Note that the four published Hipparcos distances to Blanco 1 exhibit a sizable 20\% spread (Table~\ref{table1}, see also $\alpha$ Persei).  Further research is needed, and the reader should likewise consider the  interpretations of \citet{me97}, \citet{ro99b}, \citet{so05}, and \citet{vl09,vl09b}.

\section{{\rm \footnotesize CLUSTER $\delta$ SCUTIS}}
\label{sdsct}
\subsection{{\rm \footnotesize VI PHOTOMETRY}}
The cluster $\delta$ Scutis examined are summarized in Table~\ref{table3}, along with references for their \textit{VI} photometry.  Certain sources feature $I$-band photometry not standardized to the Cousins system \citep[e.g.,][]{me67}.  Additional observations for the $\delta$ Scutis were obtained via the AAVSO's Bright Star Monitor (BSM)\footnote{http://www.aavso.org/aavsonet} and the Naval Observatory's Flagstaff Station (NOFS) (Table~\ref{table2a}).  The BSM features an SBIG ST8XME CCD (fov: $127\arcmin\times 84\arcmin$) mounted upon a 6-cm wide-field telescope located at the Astrokolkhoz telescope facility near Cloudcroft, New Mexico.  The AAVSO observations are tied to \citet{la83,la92} photometric standards according to precepts outlined in \citet{hk90} \citep[see also][]{hm08a}.  

$VI$ photometry is used since LMC and Galactic $\delta$ Scutis follow $VI$ Wesenheit relations which vary as a function of the pulsation order $n$ \citep[Fig.~\ref{fig3},][]{po10,ma10c}, thereby enabling constraints on that parameter for target $\delta$ Scutis at common or known distances (see \S \ref{dsmode}).  Furthermore, the author has advocated that RR Lyrae variables and Cepheids---which partly form the basis for the calibration used in \S \ref{swesenheit}---obey \textit{VI} Wesenheit and period-colour relations which are comparatively insensitive to metallicity \citep[][see also \citealt{bm99,ud01,bo03b,pi04,bo08}]{ma08,ma09,ma09c,ma09d,ma10,ma10b}.   For example, \citet{ma10b} reaffirmed that the slope of the \textit{VI} Wesenheit function for Milky Way Classical Cepheids \citep{be07,tu10} characterizes classical Cepheids in the LMC, NGC 6822, SMC, and IC 1613 \citep[see Fig.~2 in][]{ma10b}. Classical Cepheids in the aforementioned galaxies exhibit precise ground-based photometry, span a sizable abundance baseline, and adhere to a common \textit{VI} Wesenheit slope to within the uncertainties ($\alpha=-3.34\pm0.08(2\sigma)$, $\Delta$[Fe/H]$\simeq1$). More importantly, \citet{ma10b} noted that a negligible distance offset exists between OGLE classical Cepheids and RR Lyrae variables in the LMC, SMC, and IC 1613 as established via a $VI$ Wesenheit function, thereby precluding a dependence on metallicity.  Admittedly, the impact of a reputed metallicity effect is actively debated in the literature \citep[][and references therein]{sm04,ro08,ca09}.  By contrast there appears to be a consensus that relations which rely on $BV$ photometry are sensitive to variations in chemical abundance, and a significant break in the period-magnitude relation is apparent (\citealt{ma08,ma09c} and references therein).  The results are consistent (in part) with the findings of \S \ref{smetallicity}, however, a direct comparison is not valid.

%-------------- FIGURE 2 --------------%
\begin{figure}[!t]
\begin{center}
\includegraphics[width=5.2cm]{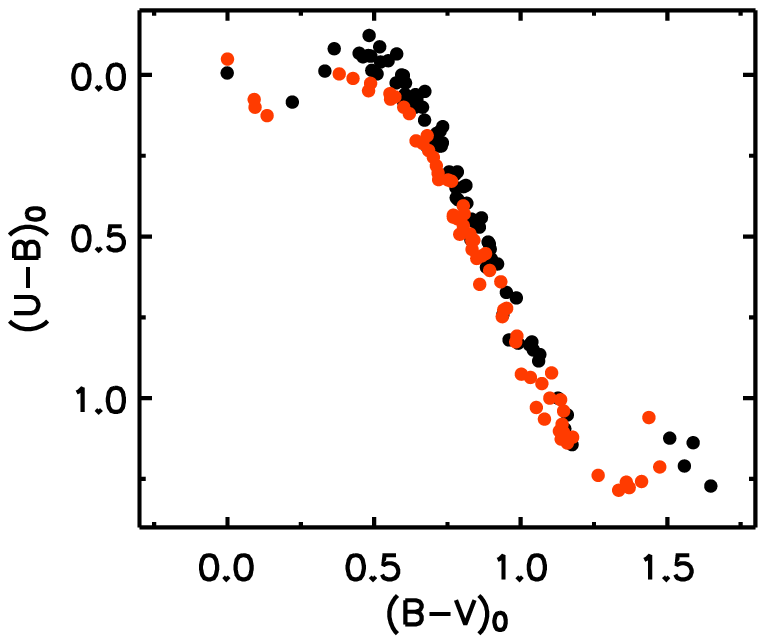} 
\includegraphics[width=5.3cm]{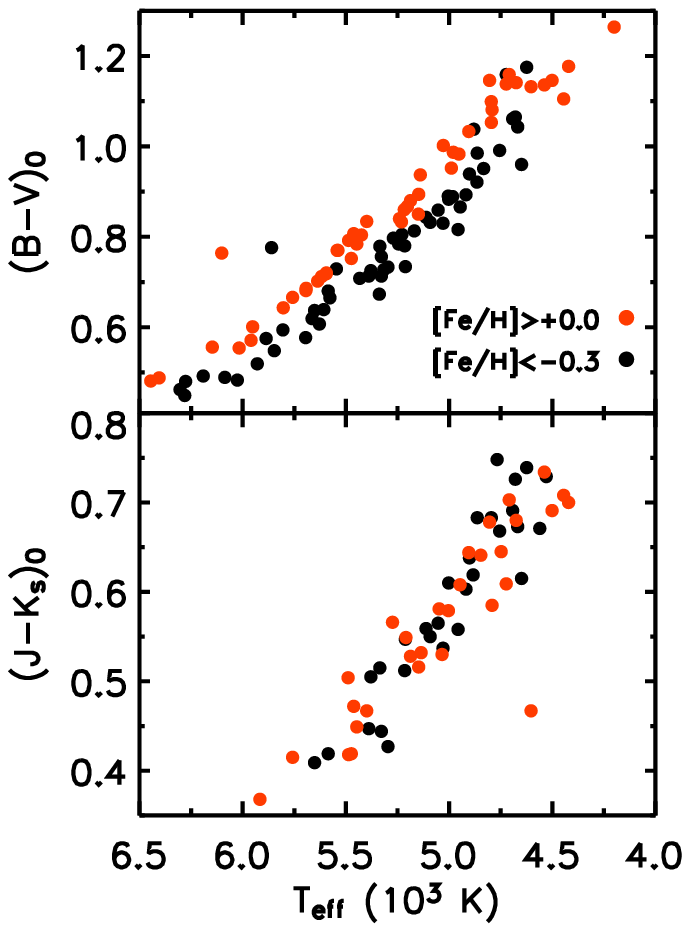} 
\caption{\small{Semi-empirical colour-$T_{e}$-[Fe/H] correlation for calibration stars (Fig.~\ref{fig1}) featured in PASTEL.  Red and black dots indicate metal-rich and metal-poor stars accordingly.  $B-V$ and $U-B$ colour indices appear sensitive to metallicity, whereas $J-K_s$ is comparatively unaffected by changes in iron abundance (see text).}}
\label{fig2}
\end{center}
\end{figure}
%-------------- FIGURE 2 --------------%

%--------------------  TABLE 2  ------------------------- %
\begin{table}[!t]
\centering
\begin{minipage}{8.5cm}
\caption[]{New Photometry for Cluster $\delta$ Scutis}
\label{table2a}
\begin{tabular}{llccc}
\hline 
\hline \rule{0pt}{2.2ex}
Star & Cluster & CMD position & \textit{V} & \textit{V--I$_c$} \\ 
        &               & (Fig.~\ref{fig2b}) &             & \\
\hline \rule{-2.5pt}{2.3ex}
EX Cnc & M67 & BS & 10.90 & 0.30 \\
EW Cnc & M67 & BS & 12.24 & 0.30 \\
HD 23156 	&	Pleiades & MS & 8.23 & 0.28 \\
HD 23194	&	Pleiades & MS & 8.08 & 0.20 \\
HD 23567 	&	Pleiades &MS & 8.29	 & 0.41\\
HD 23607 	& Pleiades & MS & 8.25	 & 0.27 \\
HD 23628	&	Pleiades & MS,BR & 7.69	& 0.24\\
HD 23643 	&	Pleiades & MS,BR & 7.79	 & 0.16 \\
\hline \\
\end{tabular}
\small{Notes: magnitudes are means of observations acquired from the AAVSO's BSM, NOFS, TASS \citep{dr06}, and \citet{me67}.  The identifiers are as follows: stars occupying the blue straggler region (BS), binary / rapid rotator sequence (BR), and main-sequence (MS) of the colour-magnitude diagram (Fig.~\ref{fig2b}).}
\end{minipage}
\end{table}
%--------------------  TABLE 2  ----------------------- %

\subsection{{\rm \footnotesize WESENHEIT MAGNITUDES}}
\label{swesenheit}
\begin{figure*}[!t]
\begin{center}
\includegraphics[width=16cm]{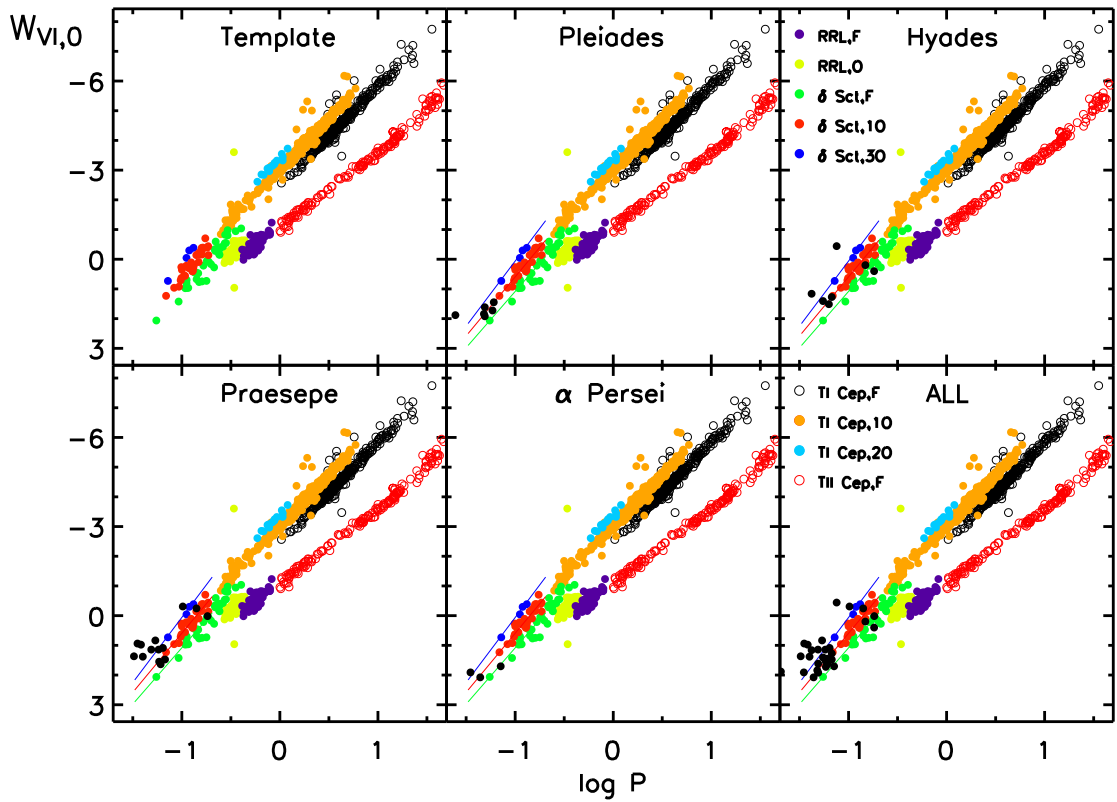} 
\caption{\small{A calibrated universal $VI$ Wesenheit template constructed from data presented in \citet{ma10c} and the latest OGLE\textit{III} observations \citep[e.g.,][]{po10}.  The cluster $\delta$ Scutis are displayed as black dots.  Wesenheit magnitudes were computed via Eqn.~\ref{eqn1} using the $JHK_s$ established cluster distances and $VI$ photometry highlighted in Tables~\ref{table1} \& \ref{table3} accordingly. First-order constraints on the inferred pulsation modes ($n$) are listed in Tables~ \ref{table2} \& \ref{table3}.}} 
\label{fig3}
\end{center}
\end{figure*}

A Wesenheit diagram segregates variables into their distinct classes (Fig.~\ref{fig3}).   Wesenheit magnitudes for the cluster $\delta$ Scutis were computed as follows:
\begin{equation}
\label{eqn1}
W_{VI,0}= \langle V \rangle -R_{VI}(\langle V \rangle - \langle I \rangle)-\mu_0 \nonumber 
\end{equation}
$\mu_0$ is the distance modulus from Table~\ref{table1} and $R_{VI}=2.55$ is the canonical extinction law, although there are concerns with adopting a colour-independent extinction law.  \textit{VI} Wesenheit magnitudes are reddening-free and comparatively insensitive to chemical composition and the width of the instability strip. The Wesenheit function is defined and discussed in the following references: \citet{ma82}, \citet{op83,op88}, \citet{mf91,mf09}, \citet{kj97}, \citet{kw01}, \citet{di04,di07}, and \citet{tu10}.

%--------------------  FIGURE 4 ------------------------%
%--------------------  FIGURE 4 ------------------------%

Cluster $\delta$ Scutis in Tables~\ref{table2} and \ref{table3} are plotted on a universal \textit{VI} Wesenheit template \citep[Fig.~\ref{fig3}, see also][]{ma10c}.  30 variables with distances measured by geometric means formed the calibration \citep[][their Table~1]{ma10c}.  The sample consisted of 8 SX Phe and $\delta$ Sct variables \citep[HIP,][]{vl07a}, 4 RR Lyrae variables \citep[HIP and HST,][]{be02b,vl07a}, 2 Type II Cepheids \citep[HIP,][]{vl07a}, and 10 classical Cepheids \citep[HST,][]{be02,be07}.  That sample was supplemented by 6 Type II Cepheids detected by \citet{ma06} in their comprehensive survey of the galaxy M106 \citep{ma09c}, which features a precise geometric-based distance estimate \citep[VLBA,][]{he99}. Type II Cepheids within the inner region of M106 were not incorporated into the calibration because of the likelihood of photometric contamination via crowding and blending \citep[\citealt{ma09c,ma10c,ma10b}, see also][]{su99,mo01,ma06,vi07,sm07}.  The stars employed were observed in the outer regions of M106 where the stellar density and surface brightness are diminished by comparison.   Admittedly, it is perhaps ironic that stars $7.2$ Mpc distant may be enlisted as calibrators owing to an absence of precise parallaxes for nearby objects.  Additional observations of new variables in M106 are forthcoming \citep{mr09}. 

LMC variables catalogued by OGLE, including the latest sample of $\delta$ Scutis \citep{po10}, were added to the Wesenheit template \citep[Fig.~\ref{fig3}, OGLE data:][see also \citealt{ud09}]{ud99,so02,so03,so08,so08b,so09}.  The LMC data were calibrated with a distance established via the geometric-anchored universal Wesenheit template \citep[$\mu_0=18.43\pm0.03 (\sigma_{\bar{x}})$,][]{ma10c}.  That distance agrees with a mean derived from 300$+$ results tabulated for the LMC at the NASA/IPAC Extragalactic Database (NED) \citep[][\footnote{http://nedwww.ipac.caltech.edu/level5/NED1D/intro.html}$^,$\footnote{http://nedwww.ipac.caltech.edu/Library/Distances/} see also Fig.~2 in \citealt{fm10}]{ms07,sm10}.  Adding \citet{tu10} list of classical Cepheids in Galactic clusters to the universal $VI$ Wesenheit calibration yields the same LMC distance with reduced uncertainties.

The universal Wesenheit template (Fig.~\ref{fig3}) unifies variables of the instability strip to mitigate uncertainties tied to establishing a distance scale based on Cepheids,  RR Lyrae, or $\delta$ Sct variables individually.   Anchoring the distance scale via the universal Wesenheit template (Fig.~\ref{fig3}) mobilizes the statistical weight of the entire variable star demographic to ensure a precise distance determination. Moreover, the universal Wesenheit template may be calibrated directly via parallaxes and apparent magnitudes, mitigating the propagation of uncertainties tied to extinction corrections. F. Benedict and coauthors are presently acquiring HST parallaxes for additional population II variables which shall bolster the template \citep{fe08}.   Further calibration could likewise ensue via variables in clusters with distances secured by dynamical means or eclipsing binaries \citep[Cluster AgeS Experiment,][]{pk04,ge06,ka07}, and variables in the Galactic bulge that are tied to a precise geometric-based distance \citep[][supported by observations from the upcoming $VVV$ survey; \citealt{mi10}]{ku03,ei05,rie09}.  

Lastly, \citet{ma10c} plotted the universal Wesenheit template (Fig.~\ref{fig3}) as a function of the fundamentalized period to highlight the \textit{first-order} $VI$ period-magnitude continuity between RR Lyrae and Type II Cepheid variables \citep[][see also \citealt{md07} and references therein]{mat06,ma09d}.  The  Wesenheit template presented here as Fig.~\ref{fig3} is plotted as a function of the dominant period, so pulsation modes may be inferred directly from the diagram. 

%--------------------  TABLE 3a  ------------------------- %
\begin{table}[!t]
\centering
\begin{minipage}{7.8cm}
\caption[]{A Comparison of $\delta$ Sct Pulsation Modes}
\label{table2}
\begin{tabular}{llccc}
\hline 
\hline \rule{0pt}{2.2ex}
Star & Cluster & $n$(W$_{VI}$) & $n$ & Source \\
\hline \rule{-2.5pt}{2.3ex}
EX Cnc & M67 & $>$3 & 3 & Z05 \\
EW Cnc & M67 & 1 or 0 & 0 & Z05 \\
HD 23156 	&	Pleiades & 1	 & 0 & F06  \\
HD 23194	&	Pleiades & $>$3	& 4 & F06  \\
HD 23567 	&	Pleiades & 1 or 2	& 0 & F06 \\
HD 23607 	& Pleiades & 1	& 0 & F06  \\
HD 23628	&	Pleiades &	1 & 0	& F06  \\
HD 23643 	&	Pleiades  & 1 or 0	& 0 & F06  \\
HD 73175	&	Praesepe &	$>$3 & 3 &  P98  \\
HD 73450	&	Praesepe &	1 & 1 &  P98 	\\
HD 73575	&	Praesepe &	$\ge$3 & 3/? & P98 \\
HD 73576	&	Praesepe &	$\ge$3 & 3	&  P98  \\
HD 73763	&	Praesepe &	$>$3 & 3/?	&  P98  \\
HD 74028	&	Praesepe &	$\ge$3 & 3	&  P98  \\
\hline \\
\end{tabular}
\small{Notes: pulsation modes (primary signal, order $n$) inferred for M67, Pleiades, and Praesepe $\delta$ Scutis from the Wesenheit template (Fig.~\ref{fig3}, $n$($W_{VI}$)) and sources in the literature ($n$). Sources are \citet[][Z05]{zh05}, \citet[][F06]{fm06}, and \citet[][P98]{pe98}.}
\end{minipage}
\end{table}
%--------------------  TABLE 3a  ----------------------- %

\subsection{{\rm \footnotesize PULSATION MODE}}
\label{dsmode}
Wesenheit ridges in Fig.~\ref{fig3} that define $\delta$ Scutis pulsating in the fundamental, first, second, and third overtone were constructed from data presented in \citet[][LMC]{po10} and \citet[][Galactic]{ma10c}.  LMC and Galactic $\delta$ Scutis pulsating in the overtone exhibit brighter $VI$ Wesenheit magnitudes ($W_{VI}$) than their fundamental mode counterparts at a given period (Fig.~\ref{fig3}, or bottom panel of Fig.~4 in \citealt{po10}).  However, a clear separation was less evident in \citet{ga10} shorter-wavelength ($VR$) observations of LMC $\delta$ Scutis (their Fig.~3), thereby motivating those authors to favour an alternate conclusion.  The present research relies on $VI$ observations of $\delta$ Scutis.

%--------------------  FIGURE 4 ------------------------%
\begin{figure*}[!t]
\begin{center}
\includegraphics[width=7cm]{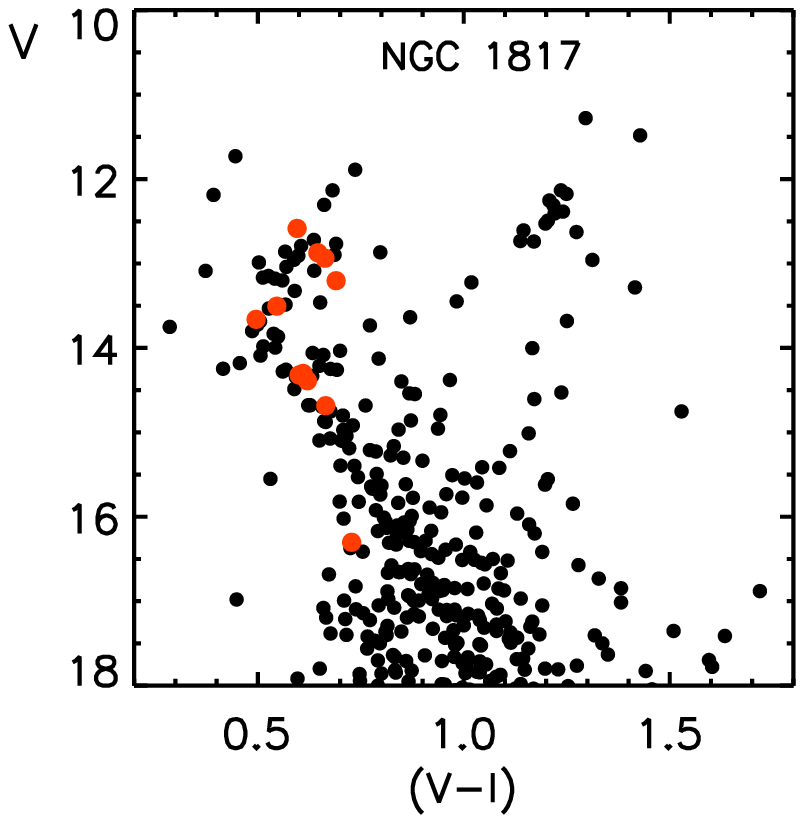} 
\includegraphics[width=7cm]{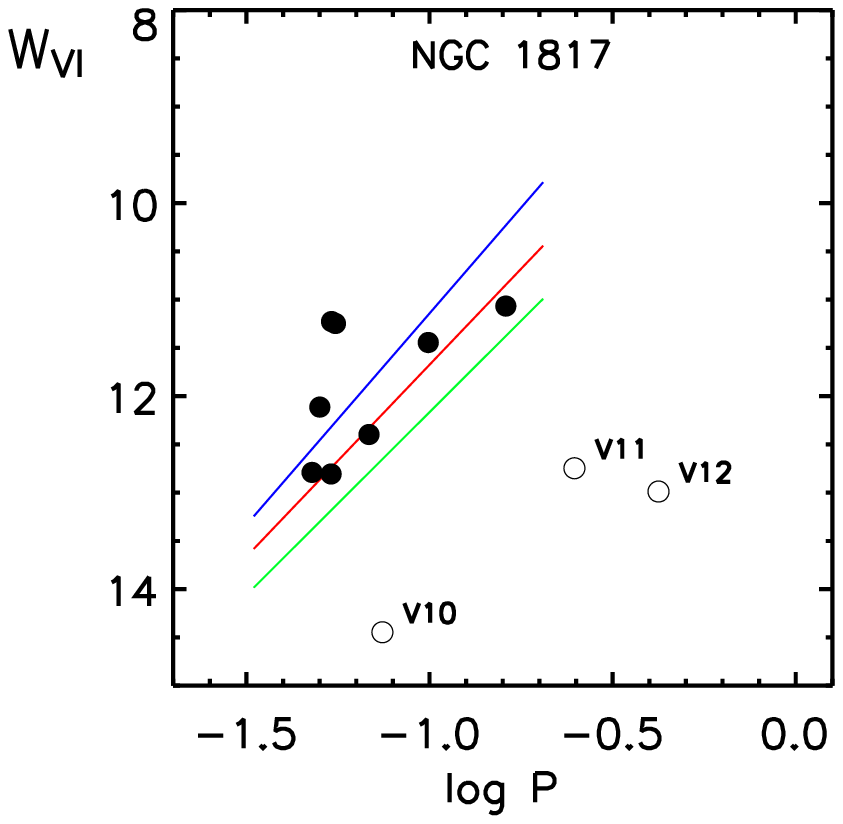} 
\caption{\small{Left, a \textit{preliminary} $VI$ colour-magnitude diagram for the open cluster NGC 1817 compiled from ARO observations. A subsample of \citealt{ar05} list of $\delta$ Scutis are highlighted by red dots.  Right, the $\delta$ Scutis are featured in a Wesenheit diagram where green, red, and blue ridges correspond to $n=0,1,3$ pulsators accordingly (right to left).  The offset from the absolute Wesenheit magnitudes (Fig.~\ref{fig3}) yields $d\simeq1.7$ kpc, assuming the variables define the $n\ge1$ boundary (see \S \ref{dsmode}).}} 
\label{fig5}
\end{center}
\end{figure*}
%--------------------  FIGURE 4 ------------------------%

Estimates of the pulsation modes (order, $n$) for the cluster $\delta$ Scutis were inferred from the Wesenheit template (Fig.~\ref{fig3}) and are summarized in Table \ref{table3}.  Pulsation modes established from Wesenheit and seismological analyses for $\delta$ Scutis in the Pleiades, Praesepe, and M67 are comparable within the mutually expected (albeit large) uncertainties ($\pm n$) (Table~\ref{table2}).  The methods identify high and low order pulsators in consistent fashion (e.g., HD23194 and HD73450, Table~\ref{table2}).  Most small-amplitude cluster $\delta$ Scutis lie on Wesenheit loci characterizing $n\ge1$ pulsators (non-fundamental mode, Fig.~\ref{fig3} and Table~\ref{table3}), and a sizable fraction are associated with $n=1$  (Table~\ref{table3}).  The results are consistent with \citet{po10} findings and past predictions \citep[][and references therein]{br00,mc07}.    

A primary source of uncertainty associated with the analysis rests with the pulsation periods adopted.  Periods for the cluster $\delta$ Scutis were taken from \citet{ro00}, the GCVS \citep{sam09}, and the AAVSO's VSX archive\footnote{http://www.aavso.org/vsx/} \citep*{wa10}.  In several instances discrepant periods are cited and newer estimates were favoured \citep[SAO 38754,][]{li05}.  Efforts to extract the primary pulsation period (high SNR) for mmag $\delta$ Scutis in $\alpha$ Persei from ARO observations were unsuccessful, likely owing to increased humidity tied to summertime observations in Nova Scotia.  That underscores the challenge such mmag pulsators present to low altitude observatories near sea-level.  An additional source of uncertainty arises from a companion's influence on the observed Wesenheit magnitudes.  The pulsation mode assigned to HD 28052, which is a spectroscopic binary and bright X-ray source, should therefore be interpreted cautiously (Table~\ref{table3}). A star's non-radial and multi-mode pulsation, and rotation/inclination along the line of sight likewise complicate a determination of $n$ solely from the pulsation period and Wesenheit magnitude.   Constraints established by Wesenheit analyses are admittedly limited by comparison to those inferred from uninterrupted space-based $\mu$mag time-series photometry (MOST, COROT, Kepler), yet the Wesenheit approach is a viable first-order tool that can be applied promptly to $\delta$ Scutis in any field and concurrently with other methods.

\subsection{{\rm \footnotesize $\delta$ SCT DISTANCE TO NGC1817}}
\label{sngc1817}
A potential role for mmag $\delta$ Scutis as distance indicators for intermediate-age open clusters is now explored using the aforementioned framework. 

\citet{ar05} observations of NGC 1817 indicated that the open cluster hosts a statistically desirable sample of 11 $\delta$ Scutis.  \citet{ba04} proper motions implied that five of those stars are not cluster members \citep[V1, V6, V8, V10, V12, see][]{ar05}.  Yet \citet{ar05} concluded that 11 variables (V1$\rightarrow$V12, excluding V10) exhibit positions in $V$ vs.~$B-V$ and $b-y$ colour-magnitude diagrams consistent with $\delta$ Sct pulsation and cluster membership.  A \textit{preliminary} $V$ vs.~$V-I$ colour-magnitude diagram for NGC 1817 (Fig.~\ref{fig5}) confirms that most variables display positions consistent with membership (except V10).  The $VI$ data were obtained from the ARO and processed using ARAP\footnote{http://www.davelane.ca/aro/arap.html} \citep{la07} and DAOPHOT\footnote{IRAF contains DAOPHOT, however a standalone newer edition can be obtained from Peter.Stetson@nrc-cnrc.gc.ca} \citep{st87}.  The ARO features an SBIG ST8XME CCD mounted upon a 35-cm telescope located near Stillwater Lake, Nova Scotia, Canada.  The ARO is a remotely operated robotic observatory \citep{la07}.  A description of the ARO observations for NGC 7062, including an analysis using VaST \citep{sl05}, shall be provided in a subsequent study.

The $VI$ Wesenheit diagram compiled for the $\delta$ Scutis (Fig.~\ref{fig5}) implies that V10, V11, and V12 either exhibit spurious data, are not \textit{bona fide} $\delta$ Scutis, or are not cluster members.  V10 is too faint, as corroborated by its position in the colour-magnitude diagram (Fig.~\ref{fig5}).  The variability detected in V11 is marred by low signal to noise and other complications \citep{ar05}.   V12 is likewise too faint ($W_{VI}$) and features a period beyond that typically expected for $\delta$ Sct variables \citep{pe07}.  An advantage to applying the Wesenheit technique is that high and low-order pulsators may be identified.  V1, V2, and V6 are likely pulsating at higher orders if the stars are members (Fig.~\ref{fig5}).  V3, V4, V7, V8, and V9 are tightly clustered and may define the $n\ge1$ boundary (Fig.~\ref{fig5}, see \S \ref{dsmode}).  The resulting distance to NGC 1817 is $d\simeq1.7$ kpc, assuming the aforementioned mode distribution.  The $\delta$ Sct distance to NGC 1817 agrees with estimates established for the cluster by other means \citep[][see references therein]{ar05}.   However, employing mmag $\delta$ Scutis as distance indicators for open clusters is complicated by the need for independently confirmed periods, and \textit{a priori} knowledge of the pulsation modes or the adoption of a mode distribution ($n\ge1$) given sizable statistics (see \S \ref{dsmode}).  Establishing the distance to NGC 1817 or the Praesepe under the latter caveat may yield a pertinent result, yet the ensuing distance to $\alpha$ Persei would be in error owing to small statistics.  $\alpha$ Persei certainly features more than 3 $\delta$ Scutis, however a detection bias emerges since nearby clusters exhibit large angular diameters which exceed the field of view of most CCDs.  Continued research is needed.

\section{{\rm \footnotesize SUMMARY AND FUTURE RESEARCH}}
\label{ssummary}
This research aimed to outline and evaluate a $VI$ Wesenheit framework for investigating cluster $\delta$ Scutis, an analysis which relied on securing absolute Wesenheit magnitudes from precise open cluster distances.  $JHK_s$ ZAMS and colour-colour relations were derived from unreddened stars near the Sun with precise Hipparcos parallaxes and were applied to infer parameters for several benchmark star clusters which host $\delta$ Scutis (Fig.~\ref{fig1}, Table~\ref{table1}).  That analysis yielded constraints on the absolute Wesenheit magnitudes ($W_{VI,0}$), evolutionary status, and pulsation modes (order, $n$) for the cluster $\delta$ Scutis (Figs.~\ref{fig2b},~\ref{fig3}, Tables~\ref{table2},~\ref{table3}).   \textit{VI} photometry for the variables were tabulated to facilitate further research (Table~\ref{table3}), and include new data acquired via the AAVSO's robotic telescopes (Table~\ref{table2a}). 

The $JHK_s$ established cluster distances are bolstered by the relative insensitivity of $J-K_s$ photometry to variations in [Fe/H] and age over the baseline examined (\S \ref{smetallicity}, Table~\ref{table1}, Fig.~\ref{fig2}).  The deep $JHK_s$ photometry extends into the low mass regime ($\simeq0.4 M_{\sun}$) and indicates that the clusters feature a common ZAMS in the infrared, and that $J-K_s$ remains constant with increasing magnitude ($M_J\gtrsim6$) for low mass M-type dwarfs whereas $J-H$ exhibits an inversion (\S \ref{smetallicity}, Figs.~\ref{fig1},~\ref{fig2b}).  The trends ensure precise ($\le5$\%) \textit{JHK$_s$} ZAMS fits by providing distinct anchor points in colour-magnitude and colour-colour diagrams (Figs.~\ref{fig1},~\ref{fig2b}).   $JHK_s$ distances for 7 of 9 clusters within 250 pc agree with \citet{vl09} revised Hipparcos estimates (Table~\ref{table1}). However, the $JHK_s$ distance to the Pleiades supports the HST estimate rather than that derived from Hipparcos data (Table~\ref{table1}). \citet{vl09,vl09b} argues in favour of the revised Hipparcos distances to open clusters and the reader is referred to that comprehensive study.  Yet the distance scale can (presently) rely on a suite of clusters that are independent of the Pleiades.  Models should be calibrated and evaluated using those nearby clusters where consensus exists regarding the distances (Table~\ref{table1}).

The general agreement between the $JHK_s$ distances derived here and \citet{vl09} Hipparcos estimates is noteworthy (Table~\ref{table1}).  The $\sim10-20$\% offset in distance for the discrepant cases (Pleiades \& Blanco 1) is not atypical for studies of open clusters \citep[Fig.~2 in][see also \citealt{di02,mp03}]{pn06}.  Several clusters feature distance estimates spanning nearly a factor of two, such as NGC 2453 \citep[][their Table 4]{ma07}, ESO 096-SC04, Collinder 419, Shorlin 1, and Berkeley 44 \citep{tu10b,tu10c}.  The universal $VI$ Wesenheit template could be applied to cluster $\delta$ Scutis so to isolate viable distance solutions, provided certain criteria are satisfied (\S \ref{sngc1817}, e.g., sizable statistics).  A Wesenheit analysis ($VI$) is a viable means for establishing pertinent constraints on a target population of $\delta$ Sct variables, particularly in tandem with other methods (Figs.~\ref{fig2b},~\ref{fig3},~\ref{fig5}, Tables~\ref{table2},~\ref{table3}).

\subsection*{{\rm \scriptsize ACKNOWLEDGEMENTS}}
\small{DJM is grateful to the following individuals and consortia whose efforts and surveys were the foundation of this study: 2MASS (R. Cutri, M. Skrutskie, S. Nikolaev), OGLE (I. Soszy{\'n}ski, R. Poleski, M. Kubiak, A. Udalski), F. van Leeuwen, F. Benedict, Z. Li, E. Michel, T. Arentoft, P. Stetson (DAOPHOT), B. Taylor, WebDA/GCPD (E. Paunzen, J.-C. Mermilliod), NOMAD (N. Zachrias), TASS (T. Droege, M. Sallman, M. Richmond), NED (I. Steer), PASTEL (C. Soubiran), Astrometry.net \citep{la10}, \citet{sk10}, CDS, AAVSO (M. Saladyga, A. Henden), arXiv, and NASA ADS. T. Krajci, J. Bedient, D. Welch, D. Starkey, A. Henden, and others kindly funded the AAVSO's BSM.} 

\begin{table*}[!t]
\centering
\begin{minipage}{10.8cm}
\caption[]{$\delta$ Scutis in Benchmark Open Clusters}
\label{table3}
\begin{tabular}{llccc}
\hline 
\hline \rule{0pt}{2.2ex}
Star & Cluster  & CMD position & $n (W_{VI})$ & \textit{VI} Photometry \\ 
     &                   & (Fig.~\ref{fig2b}) &                      & \\
\hline \rule{-2pt}{2.3ex}
SAO 38754 	&	$\alpha$ Persei	& MS	& 1&	TASS, S85 \\
HD 20919 	&	$\alpha$ Persei	& MS,BR:	& 2 or 3	&	TASS \\
HD 21553 	&	$\alpha$ Persei	&	MS & 0	&	TASS \\
HD 23156 	&	Pleiades	&	MS & 1	&	Table~\ref{table2a}\\
HD 23194	&	Pleiades	&	MS & $>$3	&	Table~\ref{table2a} \\
HD 23567 	&	Pleiades	&	MS & 1 or 2	&	Table~\ref{table2a} \\
HD 23607 	&	Pleiades	&	MS & 1	&	Table~\ref{table2a} \\
HD 23628	&	Pleiades	&	MS,BR &1	&	Table~\ref{table2a}  \\
HD 23643 	&	Pleiades	&	MS,BR &1 or 0	&	Table~\ref{table2a} \\
HD 27397	&	Hyades	&	EV & 1 or 2	&	T85  \\
HD 27459	&	Hyades	&	EV & 1	&	J06 \\
HD 27628 	&	Hyades	&	sat./EV: & 1&	 J06 \\
HD 28024 	&	Hyades	&	sat./EV: & 1 or 0	&	J06  \\
HD 28052 	&	Hyades	&	sat./EV: &0	&	J06 \\
HD 28319	&	Hyades	&	sat./EV: &$>$3	& J06	\\
HD 30780 	&	Hyades	&	sat. &$>$3	&	J06 \\
HD 73175 	&	Praesepe & MS/EV 	&	$>$3	& TASS, ME67 \\
HD 73345 	&	Praesepe	& MS/EV &	$>$3	&	TASS, ME67 \\
HD 73450 	&	Praesepe	& MS &1	&	TASS, ME67 \\
HD 73575 	&	Praesepe	&	EV &$\ge$3 &	TASS, ME67 \\
HD 73576 	&	Praesepe	& MS/EV,BR	&$\ge$3	&	TASS, ME67 \\
HD 73712	&	Praesepe	&	EV &1 or 2	&	TASS, ME67 \\
HD 73729 	&	Praesepe	& MS/EV,BR	&2 or 3	&	TASS, ME67 \\
HD 73746 	&	Praesepe	& MS	&1	&	TASS, ME67 \\
HD 73763 	&	Praesepe	& MS/EV	&$>$3	&	TASS, ME67 \\
HD 73798 	&	Praesepe	& MS/EV	&1	&	TASS, ME67 \\
HD 73819	&	Praesepe	&	EV &0 	&	TASS, ME67 \\
HD 73890	&	Praesepe	& MS/EV,BR	&$>$3	&	TASS, ME67 \\
HD 74028 	&	Praesepe	& MS/EV &$\ge$3	&	TASS, ME67 \\
HD 74050 	&	Praesepe	&	MS/EV &3	&	TASS, ME67 \\
EX Cnc & M67 & BS & $>$3	 & Table~\ref{table2a} \\
EW Cnc & M67 & BS & 1 or 0 & Table~\ref{table2a} \\
\hline \\
\end{tabular}
\small{Notes: $\delta$ Sct cluster list compiled primarily from \citet{lm99} and references therein.  The identifiers are as follows: stars occupying the blue straggler region (BS), binary / rapid rotator sequence (BR), evolved region (EV), and main-sequence (MS) of the colour-magnitude diagram (Fig.~\ref{fig2b}); Pulsation modes (primary signal, order $n$) inferred for the $\delta$ Scutis from the $VI$ Wesenheit template (Fig.~\ref{fig3}).  Hyades members may feature saturated (sat.) 2MASS photometry owing to their proximity (Table~\ref{table1}).  There are concerns regarding the photometric zero-point for bright $\delta$ Scutis sampled in the all-sky surveys \citep{hs07}.  References for the photometry are \citet[][ME67]{me67}, \citet[][T85]{tj85}, \citet[][S85]{st85}, and \citet[][J06]{j06}.}
\end{minipage}
\end{table*}


\begin{thebibliography}{150} \setlength{\itemsep}{-1.5mm}
\bibitem[Alonso et al.(1996)]{al96} Alonso, A., Arribas, S., \& Martinez-Roger, C.\ 1996, \aap, 313, 873 
\bibitem[Arentoft et al.(2005)]{ar05} Arentoft, T., Bouzid, M.~Y., Sterken, C., Freyhammer, L.~M., \& Frandsen, S.\ 2005, \pasp, 117, 601 
\bibitem[Artigau et al.(2009)]{ar09} Artigau, {\'E}., 
Bouchard, S., Doyon, R., \& Lafreni{\`e}re, D.\ 2009, \apj, 701, 1534 
\bibitem[Artigau et al.(2010)]{ar10} Artigau, {\'E}., Lamontagne, R., Doyon, R., \& Malo, L.\ 2010, \procspie, 7737,  
\bibitem[Balaguer-N{\'u}{\~n}ez et al.(2004)]{ba04} Balaguer-N{\'u}{\~n}ez, L., Jordi, C., Galad{\'{\i}}-Enr{\'{\i}}quez, D., \& Zhao, J.~L.\ 2004, \aap, 426, 819 
\bibitem[Benedict et al.(2002a)]{be02} Benedict G.~F. et al., 2002 (a), AJ, 123, 473 
\bibitem[Benedict et al.(2002b)]{be02b} Benedict, G.~F., et al.\ 2002 (b), \aj, 124, 1695 
\bibitem[Benedict et al.(2007)]{be07} Benedict G.~F. et al., 2007, AJ, 133, 1810 
\bibitem[Bonatto et al.(2008)]{bona08} Bonatto, C., Bica, E., \& Santos, J.~F.~C.\ 2008, \mnras, 386, 324 
\bibitem[Bono \& Marconi(1999)]{bm99} Bono, G., \& Marconi, M.\ 1999, New Views of the Magellanic Clouds, 190, 527 
\bibitem[Bono(2003)]{bo03b} Bono, G.\ 2003, Stellar Candles for the Extragalactic Distance Scale, 635, 85 
\bibitem[Bono et al.(2008)]{bo08} Bono, G., Caputo, F., Fiorentino, G., Marconi, M., \& Musella, I.\ 2008, \apj, 684, 102 
\bibitem[Breger(2000)]{br00} Breger, M.\ 2000, Delta Scuti and Related Stars, 210, 3
\bibitem[Catelan(2009)]{ca09} Catelan, M.\ 2009, \apss, 320, 261 
\bibitem[Cutri et al.(2003)]{cu03} Cutri, R.~M., et al.\ 2003, The IRSA 2MASS All-Sky Point Source Catalog, NASA/IPAC Infrared Science Archive.
\bibitem[Dias et al.(2002)]{di02} Dias, W.~S., Alessi, B.~S., Moitinho, A., \& L{\'e}pine, J.~R.~D.\ 2002, \aap, 389, 871 
\bibitem[Di Criscienzo et al.(2004) Di Criscienzo, Marconi \& Caputo]{di04} Di Criscienzo, M., Marconi, M., \& Caputo, F.\ 2004, \apj, 612, 1092 
\bibitem[Di Criscienzo et al.(2007)]{di07} Di Criscienzo, M., Caputo, F., Marconi, M., \& Cassisi, S.\ 2007, \aap, 471, 893 
\bibitem[Droege et al.(2006)]{dr06} Droege T.~F., Richmond M.~W., Sallman M.~P., Creager R.~P., 2006, PASP, 118, 1666
\bibitem[Eisenhauer et al.(2005)]{ei05} Eisenhauer, F., et al.\ 2005, \apj, 628, 246 
\bibitem[Feast(2008)]{fe08} Feast, M.~W.\ 2008, arXiv:0806.3019 
\bibitem[Fox Machado et al.(2006)]{fm06} Fox Machado, L., P{\'e}rez Hern{\'a}ndez, F., Su{\'a}rez, J.~C., Michel, E., \& Lebreton, Y.\ 2006, \aap, 446, 611 
\bibitem[Freedman \& Madore(2010)]{fm10} Freedman, W.~L., \& Madore, B.~F.\ 2010, arXiv:1004.1856 
\bibitem[Garg et al.(2010)]{ga10} Garg, A., et al.\ 2010, \aj, 140, 328 
\bibitem[Guinan \& Engle(2006)]{ge06} Guinan, E.~F., \& Engle, S.~G.\ 2006, \apss, 304, 5 
\bibitem[Henden \& Kaitchuck(1990)]{hk90} Henden, A.~A., \& Kaitchuck, R.~H.\ 1990, Richmond, Va.~: Willmann-Bell, c1990.
\bibitem[Henden(2002)]{he02} Henden, A.~A.\ 2002, JAAVSO, 31, 11 
\bibitem[Henden \& Sallman(2007)]{hs07} Henden, A.~A., \& Sallman, M.~P.\ 2007, The Future of Photometric, Spectrophotometric and Polarimetric Standardization, 364, 139 
\bibitem[Henden \& Munari(2008)]{hm08a} Henden, A., \& Munari, U.\ 2008, Information Bulletin on Variable Stars, 5822, 1 
\bibitem[Herrnstein et al.(1999)]{he99} Herrnstein, J.~R., et al.\ 1999, \nat, 400, 539 
\bibitem[Joner et al.(2006)]{j06} Joner, M.~D., Taylor, B.~J., Laney, C.~D., \& van Wyk, F.\ 2006, \aj, 132, 111 
\bibitem[Kaluzny et al.(2007)]{ka07} Kaluzny, J., Thompson, I.~B., Rucinski, S.~M., Pych, W., Stachowski, G., Krzeminski, W., \& Burley, G.~S.\ 2007, \aj, 134, 541 
\bibitem[Kov{\'a}cs \& Jurcsik(1997)]{kj97} Kov{\'a}cs, G., \& Jurcsik, J.\ 1997, \aap, 322, 218 
\bibitem[Kov{\'a}cs \& Walker(2001)]{kw01} Kov{\'a}cs, G., \& Walker, A.~R.\ 2001, \aap, 371, 579 
\bibitem[Kubiak \& Udalski(2003)]{ku03} Kubiak M., Udalski A., 2003, Acta Astr., 53, 117 
\bibitem[Landolt(1983)]{la83} Landolt, A.~U.\ 1983, \aj, 88, 439 
\bibitem[Landolt(1992)]{la92} Landolt, A.~U.\ 1992, \aj, 104, 340 
\bibitem[Lane(2007)]{la07} Lane D.~J., 2007, 96th Spring Meeting of the AAVSO, http://www.aavso.org/aavso/meetings/spring07present/Lane.ppt (see also http://www.davelane.ca/aro/ )
\bibitem[Lang et al.(2010)]{la10} Lang, D., Hogg, D.~W., 
Mierle, K., Blanton, M., \& Roweis, S.\ 2010, \aj, 139, 1782 
\bibitem[Li \& Michel(1999)]{lm99} Li, Z.~P., \& Michel, E.\ 1999, \aap, 344, L41 
\bibitem[Li(2005)]{li05} Li, Z.~P.\ 2005, \aj, 130, 1890 
\bibitem[Macri et al.(2006)]{ma06} Macri, L.~M., Stanek, K.~Z., Bersier, D., Greenhill, L.~J., \& Reid, M.~J.\ 2006, ApJ, 652, 1133 
\bibitem[Macri \& Riess(2009)]{mr09} Macri, L.~M., \& Riess, A.~G.\ 2009, American Institute of Physics Conference Series, 1170, 23 
\bibitem[Madore(1982)]{ma82} Madore B.~F., 1982, ApJ, 253, 575
\bibitem[Madore \& Freedman(1991)]{mf91} Madore, B.~F., \& Freedman, W.~L.\ 1991, \pasp, 103, 933 
\bibitem[Madore \& Steer(2007)]{ms07} Madore, B.~F., \& Steer, I. \ 2007, NASA/IPAC Extragalactic Database Master List of Galaxy Distances (http://nedwww.ipac.caltech.edu/level5/NED1D/intro.html)
\bibitem[Madore \& Freedman(2009)]{mf09} Madore, B.~F., \& Freedman, W.~L.\ 2009, \apj, 696, 1498 
\bibitem[Majaess et al.(2007)]{ma07} Majaess, D.~J., Turner, D.~G., \& Lane, D.~J.\ 2007, \pasp, 119, 1349 
\bibitem[Majaess et al.(2008)]{ma08} Majaess D.~J., Turner D.~G., Lane D.~J., 2008, MNRAS, 390, 1539
\bibitem[Majaess et al.(2009a)]{ma09} Majaess, D.~J., Turner, D.~G., \& Lane, D.~J.\ 2009 (a), MNRAS, 398, 263 
\bibitem[Majaess et al.(2009b)]{ma09c} Majaess, D., Turner, D., \& Lane, D.\ 2009 (b), Acta Astronomica, 59, 403 
\bibitem[Majaess(2009)]{ma09d} Majaess, D.~J.\ 2009, arXiv:0912.2928 
\bibitem[Majaess(2010a)]{ma10} Majaess, D.\ 2010 (a), Acta Astronomica, 60, 55 
\bibitem[Majaess(2010b)]{ma10b} Majaess, D.~J.\ 2010 (b), Acta Astronomica, 60, 121 
\bibitem[Majaess et al.(2010)]{ma10c} Majaess, D.~J., Turner, D.~G., Lane, D.~J., Henden, A., \& Krajci, T.\ 2010, arXiv:1007.2300 
\bibitem[Marconi \& Di Criscienzo(2007)]{md07} Marconi, M., \& Di Criscienzo, M.\ 2007, \aap, 467, 223 
\bibitem[Matsunaga et al.(2006)]{mat06} Matsunaga, N., et al.\ 2006, \mnras, 370, 1979 
\bibitem[McNamara et al.(2000)]{mc00} McNamara, D.~H., Madsen, J.~B., Barnes, J., \& Ericksen, B.~F.\ 2000, \pasp, 112, 202 
\bibitem[McNamara et al.(2007)]{mc07} McNamara, D.~H., Clementini, G., \& Marconi, M.\ 2007, \aj, 133, 2752 
\bibitem[Mendoza(1967)]{me67} Mendoza, E.~E.\ 1967, Boletin de los Observatorios Tonantzintla y Tacubaya, 4, 149 
\bibitem[Mermilliod(1991)]{me91} Mermilliod, J.-C.\ 1991, Homogeneous Means in the UBV System, \url{http://vizier.u-strasbg.fr/viz-bin/VizieR?-source=II/168}
\bibitem[Mermilliod et al.(1997)]{me97} Mermilliod, J.-C., Turon, C., Robichon, N., Arenou, F., 
\& Lebreton, Y.\ 1997, Hipparcos - Venice '97, 402, 643 
\bibitem[Mermilliod \& Paunzen(2003)]{mp03} Mermilliod, J.-C., \& Paunzen, E.\ 2003, \aap, 410, 511 
\bibitem[Minniti et al.(2010)]{mi10} Minniti, D., et al.\ 2010, New Astronomy, 15, 433 
\bibitem[Mochejska et al.(2001)]{mo01} Mochejska, B.~J., Macri, L.~M., Sasselov, D.~D., \& Stanek, K.~Z.\ 2001, arXiv:astro-ph/0103440 
\bibitem[Nikolaev et al.(2000)]{ni00} Nikolaev, S., Weinberg, M.~D., Skrutskie, M.~F., Cutri, R.~M., Wheelock, S.~L., Gizis, J.~E., \& Howard, E.~M.\ 2000, \aj, 120, 3340 
\bibitem[Opolski(1983)]{op83} Opolski A., 1983, IBVS, 2425, 1 
\bibitem[Opolski(1988)]{op88} Opolski, A.\ 1988, Acta Astronomica, 38, 375
\bibitem[Percy(2007)]{pe07} Percy, J.~R.\ 2007, Understanding variable stars / Cambridge 
University Press
\bibitem[Paunzen \& Netopil(2006)]{pn06} Paunzen, E., \& Netopil, M.\ 2006, \mnras, 371, 1641 
\bibitem[Paunzen et al.(2010)]{pa10} Paunzen, E., Heiter, U., Netopil, M., \& Soubiran, C.\ 2010, \aap, 517, A32
\bibitem[Pena et al.(1998)]{pe98} Pena, J.~H., et al.\ 1998, \aaps, 129, 9 
\bibitem[Percival et al.(2005)]{pe05} Percival, S.~M., Salaris, M., \& Groenewegen, M.~A.~T.\ 2005, \aap, 429, 887 
\bibitem[Perryman \& ESA(1997)]{pe97} Perryman, M.~A.~C., \& ESA 1997, ESA Special Publication, 1200
\bibitem[Petersen \& H{\o}g(1998)]{ph98} Petersen, J.~O., \& H{\o}g, E.\ 1998, \aap, 331, 989 
\bibitem[Pietrukowicz \& Kaluzny(2004)]{pk04} Pietrukowicz, P., \& Kaluzny, J.\ 2004, Acta Astronomica, 54, 19 
\bibitem[Pietrzy{\'n}ski et al.(2004)]{pi04} Pietrzy{\'n}ski, G., Gieren, W., Udalski, A., Bresolin, F., Kudritzki, R.-P., Soszy{\'n}ski, I., Szyma{\'n}ski, M., \& Kubiak, M.\ 2004, AJ, 128, 2815 
\bibitem[Poleski et al.(2010)]{po10} Poleski, R., et al.\ 2010, Acta Astronomica, 60, 1 
\bibitem[Reid et al.(2009)]{rie09} Reid, M.~J., Menten, K.~M., Zheng, X.~W., Brunthaler, A., \& Xu, Y.\ 2009, \apj, 705, 1548 
\bibitem[Robichon et al.(1999a)]{ro99} Robichon, N., Arenou, F., Mermilliod, J.-C., \& Turon, C.\ 1999 (a), \aap, 345, 471 
\bibitem[Robichon et al.(1999b)]{ro99b} Robichon, N., Arenou, F., Lebreton, Y., Turon, C., \& Mermilliod, J.~C.\ 1999 (b), Harmonizing Cosmic Distance Scales in a Post-HIPPARCOS Era, 167, 72 
\bibitem[Rodr{\'{\i}}guez et al.(2000)]{ro00} Rodr{\'{\i}}guez, E., L{\'o}pez-Gonz{\'a}lez, M.~J., \& L{\'o}pez de Coca, P.\ 2000, \aaps, 144, 469 
\bibitem[Romaniello et al.(2008)]{ro08} Romaniello, M., et al.\ 2008, \aap, 488, 731 
\bibitem[Samus et al.(2009a)]{sam09} Samus, N.~N., Durlevich, O.~V., \& et al.\ 2009, VizieR Online Data Catalog, 1, 2025 
\bibitem[Sarajedini et al.(2009)]{sa09} Sarajedini, A., Dotter, A., \& Kirkpatrick, A.\ 2009, \apj, 698, 1872 
\bibitem[Skrutskie et al.(2006)]{sk06} Skrutskie, M.~F., et al.\ 2006, \aj, 131, 1163
\bibitem[Skiff(2010)]{sk10} Skiff, B.\ 2010, Catalogue of Stellar Spectral Classifications, http://vizier.u-strasbg.fr/viz-bin/VizieR?-source=B/mk 
\bibitem[Smith(2004)]{sm04} Smith, H.~A.\ 2004, RR Lyrae Stars, by Horace A.~Smith, pp.~166.~ISBN 0521548179.~Cambridge, UK: Cambridge University Press, September 2004 
\bibitem[Smith et al.(2007)]{sm07} Smith, M.~C., Wo{\'z}niak, P., Mao, S., \& Sumi, T.\ 2007, \mnras, 380, 805 
\bibitem[Soderblom et al.(2005)]{so05} Soderblom, D.~R., Nelan, E., Benedict, G.~F., McArthur, B., Ramirez, I., Spiesman, W., \& Jones, B.~F.\ 2005, \aj, 129, 1616 
\bibitem[Soderblom et al.(2009)]{sod09} Soderblom, D.~R., Laskar, T., Valenti, J.~A., Stauffer, J.~R., 
\& Rebull, L.~M.\ 2009, \aj, 138, 1292 
\bibitem[Sokolovsky 
\& Lebedev(2005)]{sl05} Sokolovsky, K., \& Lebedev, A.\ 2005, 12th Young Scientists' Conference on Astronomy and Space Physics, 79 
\bibitem[Soszy{\'n}ski et al.(2002)]{so02} Soszy{\'n}ski, I., et al.\ 2002, Acta Astronomica, 52, 369 
\bibitem[Soszy{\'n}ski et al.(2003)]{so03} Soszy{\'n}ski, I., et al.\ 2003, Acta Astronomica, 53, 93 
\bibitem[Soszy{\'n}ski et al.(2008)]{so08} Soszy{\'n}ski, I., et al.\ 2008, Acta Astronomica, 58, 293 
\bibitem[Soszy{\'n}ski et al.(2008b)]{so08b} Soszy{\'n}ski, I., et al.\ 2008 (b), Acta Astronomica, 58, 163 
\bibitem[Soszy{\'n}ski et al.(2009)]{so09} Soszy{\'n}ski, I., et al.\ 2009, Acta Astronomica, 59, 1 
\bibitem[Soubiran et al.(2010)]{sou10} Soubiran, C., Le Campion, J.-F., Cayrel de Strobel, G., \& Caillo, A.\ 2010, \aap, 515, A111 
\bibitem[Stanek \& Udalski(1999)]{su99} Stanek, K.~Z., \& Udalski, A.\ 1999, arXiv:astro-ph/9909346 
\bibitem[Stauffer et al.(1985)]{st85} Stauffer, J.~R., Hartmann, L.~W., Burnham, J.~N., \& Jones, B.~F.\ 1985, \apj, 289, 247 
\bibitem[Stauffer et al.(2003)]{st03} Stauffer, J.~R., Jones, B.~F., Backman, D., Hartmann, L.~W., Barrado y Navascu{\'e}s, D., Pinsonneault, M.~H., Terndrup, D.~M., \& Muench, A.~A.\ 2003, \aj, 126, 833 
\bibitem[Steer \& Madore(2010)]{sm10} Steer, I. \& Madore, B.~F. \ 2010, NED-D: A Master List of Redshift-Independent Extragalactic Distances (http://nedwww.ipac.caltech.edu/Library/Distances/)
\bibitem[Stetson(1987)]{st87} Stetson, P.~B.\ 1987, \pasp, 99, 191 
\bibitem[Strai{\v z}ys \& Laugalys(2009)]{sl09} Strai{\v z}ys, V., \& Laugalys, V.\ 2009, Baltic Astronomy, 18, 141 
\bibitem[Taylor \& Joner(1985)]{tj85} Taylor, B.~J., \& Joner, M.~D.\ 1985, \aj, 90, 479 
\bibitem[Taylor(2008)]{ta08} Taylor, B.~J.\ 2008, \aj, 136, 1388 
\bibitem[Templeton(2009)]{te09} Templeton, M.~R.\ 2009, Astronomical Society of the Pacific Conference Series, 412, 187 
\bibitem[Turner(1979)]{tu79} Turner, D.~G.\ 1979, \pasp, 91, 
642 
\bibitem[Turner et al.(2009a)]{tu09} Turner, D.~G., Majaess, D.~J., Lane, D.~J., Szabados, L., Kovtyukh, V.~V., Usenko, I.~A., \& Berdnikov, L.~N.\ 2009, American Institute of Physics Conference Series, 1170, 108 
\bibitem[Turner(2010a)]{tu10} Turner, D.~G.\ 2010 (a), \apss, 326, 219 
\bibitem[Turner(2010b)]{tu10b} Turner, D.~G.\ 2011 (b), arXiv:1102.0347
\bibitem[Turner(2010c)]{tu10c} Turner, D.~G.\ 2010 (c), \textit{submitted.}
\bibitem[Udalski et al.(1999)]{ud99} Udalski A. et al., 1999, Acta Astr., 49, 223
\bibitem[Udalski et al.(2001)]{ud01} Udalski, A., Wyrzykowski, L., Pietrzynski, G., Szewczyk, O., Szymanski, M., Kubiak, M., Soszy{\'n}ski, I., \& Zebrun, K.\ 2001, Acta Astronomica, 51, 221 
\bibitem[Udalski(2009)]{ud09} Udalski, A.\ 2009, Astronomical Society of the Pacific Conference Series, 403, 110 
\bibitem[van Altena et al.(1997)]{va97} van Altena, W.~F., et al.\ 1997, \apjl, 486, L123 
\bibitem[van 
Leeuwen(1999)]{vl99} van Leeuwen, F.\ 1999, \aap, 341, L71 
\bibitem[van Leeuwen(2007)]{vl07a} van Leeuwen, F.\ 2007, \aap, 474, 653 
\bibitem[van Leeuwen(2009a)]{vl09} van Leeuwen, F.\ 2009 (a), \aap, 497, 209 
\bibitem[van Leeuwen(2009b)]{vl09b} van Leeuwen, F.\ 2009 (b), \aap, 500, 505 
\bibitem[Vilardell et al.(2007)]{vi07} Vilardell, F., Jordi, C., \& Ribas, I.\ 2007, \aap, 473, 847 
\bibitem[Watson et al.(2010)Watson, Henden \& Price]{wa10} Watson, C., Henden, A.~A., \& Price, A.\ 2010, VizieR Online Data Catalog, 1, 2027 
\bibitem[Zacharias et al.(2004)]{za04} Zacharias, N., Monet, D.~G., Levine, S.~E., Urban, S.~E., Gaume, R., 
\& Wycoff, G.~L.\ 2004, Bulletin of the American Astronomical Society, 36, 1418 
\bibitem[Zhang et al.(2005)]{zh05} Zhang, X.-B., Zhang, R.-X., \& Li, Z.-P.\ 2005, \cjaa, 5, 579 
\end{thebibliography}
\end{document}